\documentclass[conference]{IEEEtran}

\IEEEoverridecommandlockouts

\usepackage{graphicx}
\usepackage{cite}
\usepackage{url}
\usepackage[cmex10]{amsmath}
\usepackage{amssymb}
\usepackage{algorithm}
\usepackage{algpseudocode}
\usepackage{cases}
\usepackage[caption=false,font=footnotesize]{subfig}
\usepackage{color}
\usepackage{cite}
\usepackage{epstopdf}
\usepackage{calc}
\usepackage{array}

\DeclareGraphicsExtensions{.pdf,.eps,.png,.jpg,.mps}

\begin{document}
\title{Cyclic Weighted Centroid Localization for Spectrally Overlapped Sources in Cognitive Radio Networks}
\author{\IEEEauthorblockN{Shailesh Chaudhari and Danijela Cabric}
\IEEEauthorblockA{Department of Electrical Engineering\\
University of California, Los Angeles\\
Email: schaudhari@ucla.edu, danijela@ee.ucla.edu}
\thanks{This work has been supported by the National Science Foundation under CNS grant 1117600.}}
\maketitle


\begin{abstract}
We consider the problem of localizing spectrally overlapped sources in cognitive radio networks. A new weighted centroid localization algorithm (WCL) called Cyclic WCL is proposed, which exploits the cyclostationary feature of the target signal to estimate its location coordinates. In order to analyze the algorithm in terms of root-mean-square error (RMSE), we model the location estimates as the ratios of quadratic forms in a Gaussian random vector. With analysis and simulation, we show the impact of the interferer location and its modulation scheme on the RMSE. We also study the RMSE performance of the algorithm for different power levels of the target and the interference. Further, the comparison between Cyclic WCL and WCL w/o cyclostationarity is presented.
It is observed that the Cyclic WCL provides significant performance gain over WCL.
\end{abstract}

\IEEEpeerreviewmaketitle



\section{Introduction}
\label{sec:Introduction}
Transmitter location is a key piece of information in Cognitive Radio (CR) networks. Many advanced algorithms for object tracking, location-aware routing, and power management require the knowledge of the transmitter location. This information can also enable spectrum policy enforcement and improved spectrum sensing in CR networks.

Localization of Primary Users (PUs) in a CR network presents unique challenges because the PUs do not cooperate with the radios in the localization process. Therefore, traditional methods based on time-of-arrival (TOA) and time-delay-of-arrival (TDOA), such as \cite{Kaune2012}, are not applicable in this case. The weighted centroid localization (WCL) algorithm, which does not require any cooperation from PUs, has been analyzed in \cite{Wang2011}. This algorithm relies on the received signal power at different receivers in the network. However, in the presence of spectral interference, the received power is the sum of the powers received from both the target and the interferer. This results in shifting of the target location estimates away from the target. Therefore, if there is any interferer in the network, the WCL algorithm yields higher localization error. We can argue that any algorithm based on Received Signal Power or Received Signal Strength Indicator (RSSI) will suffer the same fate.

One way to tackle this problem is to use a distributed antenna system as proposed in \cite{Cabric2012}, which localizes multiple transmitters in the network. This method uses the MUltiple SIgnal Classification (MUSIC) technique based on spatially steered covariance matrix as proposed in \cite{Krolik1989}. However, as with any MUSIC-based method, this algorithm, too, requires prohibitive grid search on the area in order to localize the transmitters.

In this paper, we propose a low complexity cyclostationary-based WCL algorithm, called Cyclic Weighted Centroid Localization (Cyclic WCL) to locate spectrally overlapped sources. Similar to the WCL in \cite{Wang2011}, the Cyclic WCL also directly provides the x-y coordinate estimates of the target without grid search.

We consider a CR network that includes a target signal with cyclic frequency, say $\alpha_t$, and an interferer that does not have any cyclostationary feature at this cyclic frequency. We exploit the fact that the Cyclic Auto-Correlation (CAC) of the received signal, computed at $\alpha_t$, at any radio in the network is proportional to the power transmitted by the target. We use the strength of the CAC to compute weights that are used in Cyclic WCL to estimate the x-y coordinates of the target. We present detailed system model and algorithm in Section \ref{sec:Model}.

Further, in order to analyze the proposed algorithm, we show that the location estimates can be reduced to \textit{ratios of quadratic forms in a Gaussian random vector}. Using this form, we show how to compute the root-mean-square error (RMSE) for the proposed estimator. This analysis is presented in Section \ref{sec:Theoretical}. In Section \ref{sec:Results}, we discuss simulation results and show how RMSE performance is affected by the location of the interferer, its modulation scheme and the target and the interferer power levels. We also show, in this Section, that Cyclic WCL outperforms WCL in the presence of an interferer. Finally, Section \ref{sec:Conclusion} concludes the paper.

\section{System Model and Algorithm}
\label{sec:Model}
\begin{figure*}
\centering
\includegraphics[width=\columnwidth]{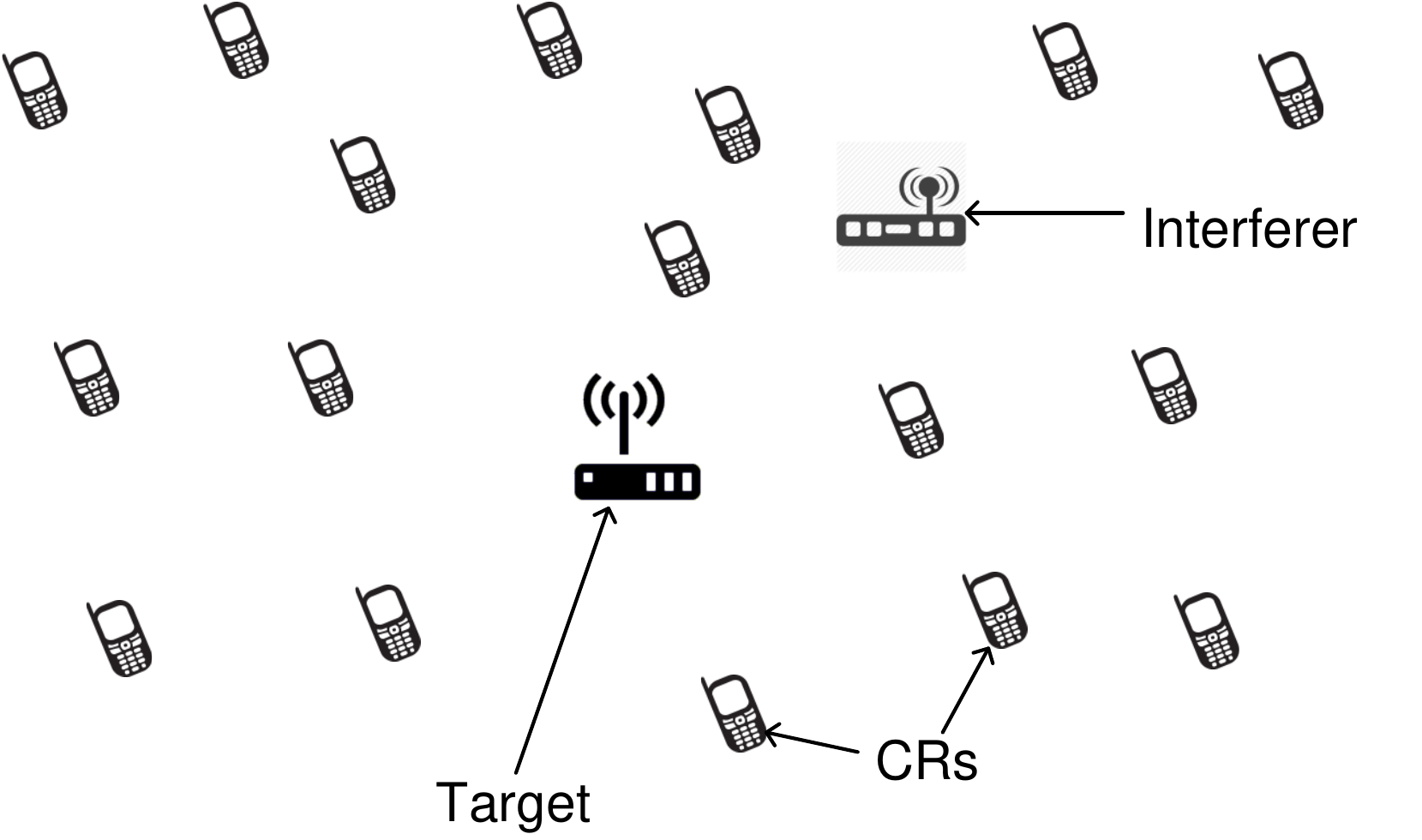}
\caption{System Schematic: Cognitive radio network with a target and a spectrally overlapped interference.}
\label{fig:system_schematic}
\end{figure*}

Let us consider a target-centric CR network in which the target PU to be localized is at the center and the interferer location is $\boldsymbol{L_{i}}=[x_{i},y_{i}]^{T}$. We denote the location coordinates of the target as $\boldsymbol{L_{t}}=[x_{t},y_{t}]^{T}=[0,0]^{T}$. There are $K$ radios in the network independently located in both $x$ and $y$ dimensions. Their coordinates are known and are denoted as $\boldsymbol{L_{k}}=[x_{k},y_{k}]^{T}, 1\leq k\leq K$. The schematic of the system is shown in Fig.~\ref{fig:system_schematic}.

Let $s_t$ be the signal transmitted by the target and $s_i$ be the signal transmitted by the interferer. The target signal, $s_t$ is a cyclostationary signal with known cyclic frequency $\alpha_t$. The interferer signal, $s_i$ may occupy the same band as the target signal, and it may be cyclostationary. However its cyclic frequency is not the same as that of the target signal. The target signal, $s_t$, and the interferer signal, $s_i$, are normalized to have unit power.

The powers received at the $k^{th}$ CR from the target ($p_{t,k}$) and from the interferer ($p_{i,k}$) are modeled using the simplified path-loss formula as follows:
\begin{equation}
p_{t,k}=p_{t}\left(\frac{||\boldsymbol{L_{t}}-\boldsymbol{L_{k}}||}{d_{0}}\right)^{-\gamma}\label{eq:ptk}
\end{equation}
\begin{equation}
p_{i,k}=p_{i}\left(\frac{||\boldsymbol{L_{i}}-\boldsymbol{L_{k}}||}{d_{0}}\right)^{-\gamma}\label{eq:pik}
\end{equation}
where $p_t$ and $p_i$ are the power levels at reference distance $d_0$ from the target and the interferer, and $\gamma$ is the path-loss exponent. We assume that the noise power at each node in the network is the same and is denoted by $\sigma^2_w$. Let $w_k\sim N(0,\sigma^2_w)$ denote AWGN noise at the $k^{th}$ CR. The received signal at the $k^{th}$ radio can then be written as
\begin{equation}
r_k(n)=\sqrt{p_{t,k}} s_t(n)+\sqrt{p_{i,k}} s_i(n)+w_k(n).
\label{eq:rk}
\end{equation}
Here $r_k(n)$ is the $n^{th}$ sample of received signal $r_k$. In Cyclic WCL algorithm, we use the strength of the CAC of the received signal, $r_k$ to compute weights, $\eta_k$, for each node. First, we estimate the CAC of the received signal at cyclic frequency $\alpha_t$ using $N$ samples of the received signal observed at sampling rate $1/T_s$. 

\begin{equation}
{\hat R_{{r_k}}} = \frac{1}
{N}\mathop \sum \limits_{n = 1}^N |{r_k}(n){|^2}{e^{ - j2\pi {\alpha _t}n{T_s}}}
\label{eq:cac_r}
\end{equation}

Using CAC estimates, we obtain weights ${\eta _k} = \frac{{|{\hat R_{{r_k}}}{|^2}}}
{|{{\hat R_{{r_k}}}{|^2}_{max}}}$, where 
${|{\hat R_{{r_k}}}{|^2}_{max}} $ = max\{${|{\hat R_{{r_k}}}{|^2}, 1\leq k \leq K}$\}. The coordinates of the target are estimated as follows:
\begin{equation}
\boldsymbol{{\hat L}_t} = \frac{{\sum\limits_{k = 1}^K {{\eta _k}} \boldsymbol{L_k}}}
{{\sum\limits_{k = 1}^K {{\eta _k}} }} = \frac{{\sum\limits_{k = 1}^K {|{\hat R_{{r_k}}}{|^2}} \boldsymbol{L_k}}}
{{\sum\limits_{k = 1}^K {|{\hat R_{{r_k}}}{|^2}} }}
\label{eq:lt}
\end{equation}

In the next section, we present the theoretical analysis of the proposed estimator in terms of root-mean-square error (RMSE).
\section{Theoretical Analysis}
\label{sec:Theoretical}
\subsection{Cyclic WCL estimates as ratios of quadratic forms of a vector}
In this section, we formulate the estimates of x-y coordinates as ratios of quadratic forms of a vector. Later, we show that this vector is, in fact, a Gaussian random vector. First, we define Cyclic Auto-Correlation (CAC) of a signal $u(n)$ and Cyclic Cross-Correlation (CCC) between signals, $u(n)$ and $v(n)$, at cyclic frequency $\alpha_t$ as:
\begin{equation}
\begin{gathered}
  {{\hat R}_u} = \frac{1}
{N}\mathop \sum \limits_{n = 1}^N |u(n){|^2}{e^{ - j2\pi {\alpha _t}n{T_s}}} \hfill \\
  {{\hat R}_{uv}} = \frac{1}
{N}\mathop \sum \limits_{n = 1}^N 2\operatorname{Re} \{ u(n)v{(n)^*}\} {e^{ - j2\pi {\alpha _t}n{T_s}}}. \hfill \\ 
\end{gathered} 
\end{equation}
In the above definition and the rest of the analysis, $(\hat{\text{       }})$ indicates the estimate based on $N$ samples, and $(^*)$ indicates complex conjugate. Using the above definition, we can decompose the CAC of the received signal in (\ref{eq:cac_r}) as 
\begin{align}
\nonumber
{{\hat R}_{{r_k}}} ={\text{       }} &{p_{t,k}}{{\hat R}_{{s_t}}} + {p_{i,k}}{{\hat R}_{{s_i}}} + \sqrt {{p_{t,k}}{p_{i,k}}} {{\hat R}_{{s_t}{s_i}}} \\&+ {{\hat R}_{{w_k}}}  
  {\text{       }} + \sqrt {{p_{t,k}}} {{\hat R}_{{s_t}{w_k}}} + \sqrt {{p_{i,k}}} {{\hat R}_{{s_i}{w_k}}}.   
\label{eq:cac_r_expanded}
\end{align}
In the above equation, it can be shown that mean and variance of the CAC of the noise process at cyclic frequency $\alpha_t$ are very small when number of samples, $N$, is sufficiently large. Further, crosscorrelation between the noise $w_k$ and the signal $s_t$ is projection of the signal on noise subspace. Since the signal is orthogonal to the noise subspace, this crosscorrelation is negligible. The Cyclic Cross-Correlation ($\hat R_{s_tw_k}$) between $s_t$ and  $w_k$, projects the crosscorrelation on the complex exponentials $e^{-j2\pi\alpha_t n T_s}$. However, because the crosscorrelation itself is negligible, we can ignore its projection on the complex exponentials for the purpose of analysis. Similar argument is valid for $\hat R_{s_iw_k}$. Therefore, last three terms in RHS of (\ref{eq:cac_r_expanded}) are eliminated and we have a simplified form:
\begin{equation}
{{\hat R}_{{r_k}}} = {p_{t,k}}{{\hat R}_{{s_t}}} + {p_{i,k}}{{\hat R}_{{s_i}}} + \sqrt {{p_{t,k}}{p_{i,k}}} {{\hat R}_{{s_t}{s_i}}}.
\label{eq:cac_r_simplified}
\end{equation}
Next, let us define three 3x1 vectors: $\boldsymbol{\hat \theta_r} , \boldsymbol{\hat \theta_i}$ which contain real and imaginary parts of the CAC and the CCC and $\boldsymbol{p_k}$, which contains power received at the $k^{th}$ receiver from the target and the interferer.
\begin{equation}
\begin{gathered}
\boldsymbol{\hat \theta_r} = {\left[ {\operatorname{Re} \{ {{\hat R}_{{s_t}}}\} {\text{ }}\operatorname{Re} \{ {{\hat R}_{{s_i}}}\} {\text{ }}\operatorname{Re} \{ {{\hat R}_{{s_t}{s_i}}}\}} \right]^T} \hfill \\
\boldsymbol{{\hat \theta_i }} = {\left[ {\operatorname{Im} \{ {{\hat R}_{{s_t}}}\} {\text{ }}\operatorname{Im} \{ {{\hat R}_{{s_i}}}\} {\text{ }}\operatorname{Im} \{ {{\hat R}_{{s_t}{s_i}}}\} } \right]^T} \hfill \\
\boldsymbol{p_k} = {\left[ {{p_{t,k}}{\text{ }}{p_{i,k}}{\text{ }}\sqrt {{p_{t,k}}{p_{i,k}}}} \right]^T} .\hfill \\ 
\label{eq:vector_def}
\end{gathered} 
\end{equation}
Now we can write the estimate of the x-coordinate of the target $x_t$ in terms of the ratio of weighted sum of a vector norm:
\begin{equation}
{{\hat x}_t} = \frac{{\sum\limits_{k = 1}^K {|{{\hat R}_{{r_k}}}{|^2}} {x_k}}}
{{\sum\limits_{k = 1}^K {|{{\hat R}_{{r_k}}}{|^2}} }} = \frac{{{{\sum\limits_{k = 1}^K {\left\| {\left[ \begin{gathered}
  {\boldsymbol{\hat \theta_r }^T} \hfill \\
  {\boldsymbol{\hat \theta_i }^T} \hfill \\ 
\end{gathered}  \right]\boldsymbol{p_k}} \right\|} }^2}{x_k}}}
{{\sum\limits_{k = 1}^K {{{\left\| {\left[ \begin{gathered}
  {\boldsymbol{\hat \theta_r }^T} \hfill \\
  {\boldsymbol{\hat \theta_i }^T} \hfill \\ 
\end{gathered}  \right]\boldsymbol{p_k}} \right\|}^2}} }}
\label{eq:xt}
\end{equation}
Further, we define the symmetric matrices $\boldsymbol{A_{p,x}}$ and $\boldsymbol{A_{x}}$ and the positive definite matrices $\boldsymbol{B_{p}}$ and $\boldsymbol{B}$:
\begin{equation}
\begin{gathered}
  \boldsymbol{A_{p,x}} = \mathop \sum \limits_{k = 1}^K \boldsymbol{p_k}{\text{ }}{x_k}\boldsymbol{p_k}^T \hfill \\
  \boldsymbol{A_x} = diag(\boldsymbol{A_{p,x}},\boldsymbol{A_{p,x}}) \hfill \\
  \boldsymbol{B_p} = \mathop \sum \limits_{k = 1}^K  \boldsymbol{p_k}\boldsymbol{p_k}^T \hfill \\
  \boldsymbol{B} = diag(\boldsymbol{B_p},\boldsymbol{B_p}). \hfill \\ 
\end{gathered}
\label{eq:matrix_def}
\end{equation}
Accodring to the above definition, $\boldsymbol{A_{p,x}}$ is the sum of the weighted outer product of vectors $\boldsymbol{p_k}, (1\leq k\leq K)$. Hence $\boldsymbol{A_{p,x}}$ and $\boldsymbol{A_{x}}$ are symmetric matrices. The proof of positive definiteness of $\boldsymbol{B_p}$ and $\boldsymbol{B}$ is presented in the Appendix.

As shown in the Appendix, $\hat x_t$ can be written in terms of the above matrices and a 6x1 vector $\boldsymbol{\hat \theta}  = {\left[ {{\boldsymbol{\hat \theta_r }^T}{\text{ }}{\boldsymbol{\hat \theta_i }^T}} \right]^T}$ : 
\begin{equation}
{{\hat x}_t} = \frac{{{\boldsymbol{\hat \theta }^T}\boldsymbol{A_x}\boldsymbol{\hat \theta} }}
{{{\boldsymbol{\hat \theta }^T}\boldsymbol{B}\boldsymbol{\hat \theta} }}.
\label{eq:xt_2}
\end{equation}

Similarly, the estimate of the y-coordinate can be written as:
\begin{equation}
{{\hat y}_t} = \frac{{{\boldsymbol{\hat \theta }^T}\boldsymbol{A_y}\boldsymbol{\hat \theta} }}
{{{\boldsymbol{\hat \theta }^T}\boldsymbol{B}\boldsymbol{\hat \theta} }},
\label{eq:yt_2}
\end{equation}
where $\boldsymbol{A_y}$ is obtained by replacing $x_k$'s with $y_k$'s in (\ref{eq:matrix_def}).

\subsection{RMSE Computation}
We observe that the elements of vector $\boldsymbol{\hat{\theta}}$ are estimates of Cyclic-Auto-Correlations $\hat{R}_{s_t}$, $\hat{R}_{s_i}$ and Cyclic-Cross-Correlation $\hat{R}_{s_t,s_i}$ based on $N$ samples of the signals $\hat s_t$ and $\hat s_i$. Here we invoke \cite[Thm. 1]{Dandawate1994}, which proves that, for sufficiently large $N$, these estimates can be modeled as Gaussian Random variables. The required condition on the signals in this theorem is as follows: for the processes $s_i$ and $s_t$, the samples that are well separated in time should be approximately independent. This assumption holds true for any digital communication signal. Therefore, individual elements of $\boldsymbol{\hat{\theta}}$ are real Gaussian variables. Hence, we model $\boldsymbol{\hat{\theta}}$ as a Gaussian Random Vector with mean $E[\boldsymbol{\hat{\theta}}]=\boldsymbol{\theta}$ and covariance matrix  $\boldsymbol{\Sigma_\theta}$ i.e. $\boldsymbol{\hat{\theta}} \sim N(\boldsymbol{\theta},\boldsymbol{\Sigma_\theta})$.

Now we can argue that the estimates of x and y coordinates of the target, as derived in (\ref{eq:xt_2}) and (\ref{eq:yt_2}), are \textit{ratios of quadratic forms of a Gaussian vector} $\boldsymbol{\hat{\theta}}$. These ratios are well studied in statistics. Algorithms to compute moments of such ratios have been derived in \cite{Bao2013}. In our model, because the target is located at $[0,0]$, mean square error (MSE) is given as $E[\hat{x}_t^2]+E[\hat{y}_t^2]$. Hence we find second-order moments of $\hat{x}_t$ and $\hat{y}_t$.

Proposition 1 in \cite{Bao2013} provides the necessary and sufficient condition for the existence of $E[\hat{x}_t^2]$ and $E[\hat{y}_t^2]$. This condition essentially says that if matrix $\boldsymbol{B}$ is positive definite, the second-order moments always exist. We have shown in the Appendix that $\boldsymbol{B}$ is positive definite. Therefore we can use \cite[Eqn. 10]{Bao2013} to find these moments. We used the MATLAB code available at \cite{Web1999} to obtain moments of ratio of quadratic forms in Gaussian variable. Root-mean-square error is then calculated by taking square-root of $E[\hat{x}_t^2]+E[\hat{y}_t^2]$.

\section{Simulation Results and Discussion}
\label{sec:Results}
In this section, we evaluate the RMSE performance of the proposed algorithm in four different scenarios. We study the impact of different positions of the interferer and its modulation schemes on the location estimation error. We also examine how different power levels of the target affect the RMSE. Finally, we compare Cyclic WCL with WCL to show that our algorithm performs better when there is an interferer in the network.

In the simulations, the unit for all distance and coordinates is $meter$. We consider a CR network distributed over a square shaped area of 100 x 100 with x- and y-coordinated in range [-50,50] i.e. $-50 \leq x_k \leq 50$ and $-50 \leq y_k \leq 50$, where $x_k$ and $y_k$ are location coordinates of the $k^{th}$ radio. As mentioned before, the target is located at the center i.e. $[x_t,y_t]$ = $[0,0]$, the interferer is located at $[x_i, y_i]$. The remaining simulation parameters are listed in Table \ref{tab:table}.

\begin{table}
\caption{Simulation Parameters}
\centering
\begin{tabular}{|m{3cm}|c|}
\hline 
Parameter & Value\tabularnewline
\hline 
\hline 
Target Location & $[x_{t,}y_{t}]=[0,0]$\tabularnewline
\hline 
Interferer Location & $[x_{i,}y_{i}]=\{[10,10],[20,20],[30,30]\}$\tabularnewline
\hline 
CR Location & $[x_{k,}y_{k}];-50\leq x_{k}\leq50,-50\leq y_{k}\leq50$\tabularnewline
\hline 
Target Power & $p_{t}=\{10, 20, 30\}dBm$\tabularnewline
\hline 
Interferer Power & $p_{i}=\{15,20,25,30,35,40\}dBm$\tabularnewline
\hline 
Noise PSD & $N_0=-174dBm/Hz$\tabularnewline
\hline 
Reference Distance & $d_{0}=1$\tabularnewline
\hline 
Path-loss exponent & $\gamma=3.8$\tabularnewline
\hline 
Target and interferer carrier frequency & $f_{c}=2.4GHz$\tabularnewline
\hline 
Bandwidth of the target & $20MHz$\tabularnewline
\hline 
Bandwidth of the interferer & $40MHz$\tabularnewline
\hline 
Cyclic Frequency  of the target & $\alpha_{t}=20MHz$\tabularnewline
\hline 
Cyclic Frequency of the interferer & $\alpha_{i}=40MHz$\tabularnewline
\hline 
Sampling frequency & $f_{s}=200MHz$\tabularnewline
\hline 
Number of samples & $N=5000$\tabularnewline
\hline \hline
\end{tabular}
\label{tab:table}
\end{table}
\subsection{Position of the interferer}
\begin{figure}
\centering
\includegraphics[width=\columnwidth]{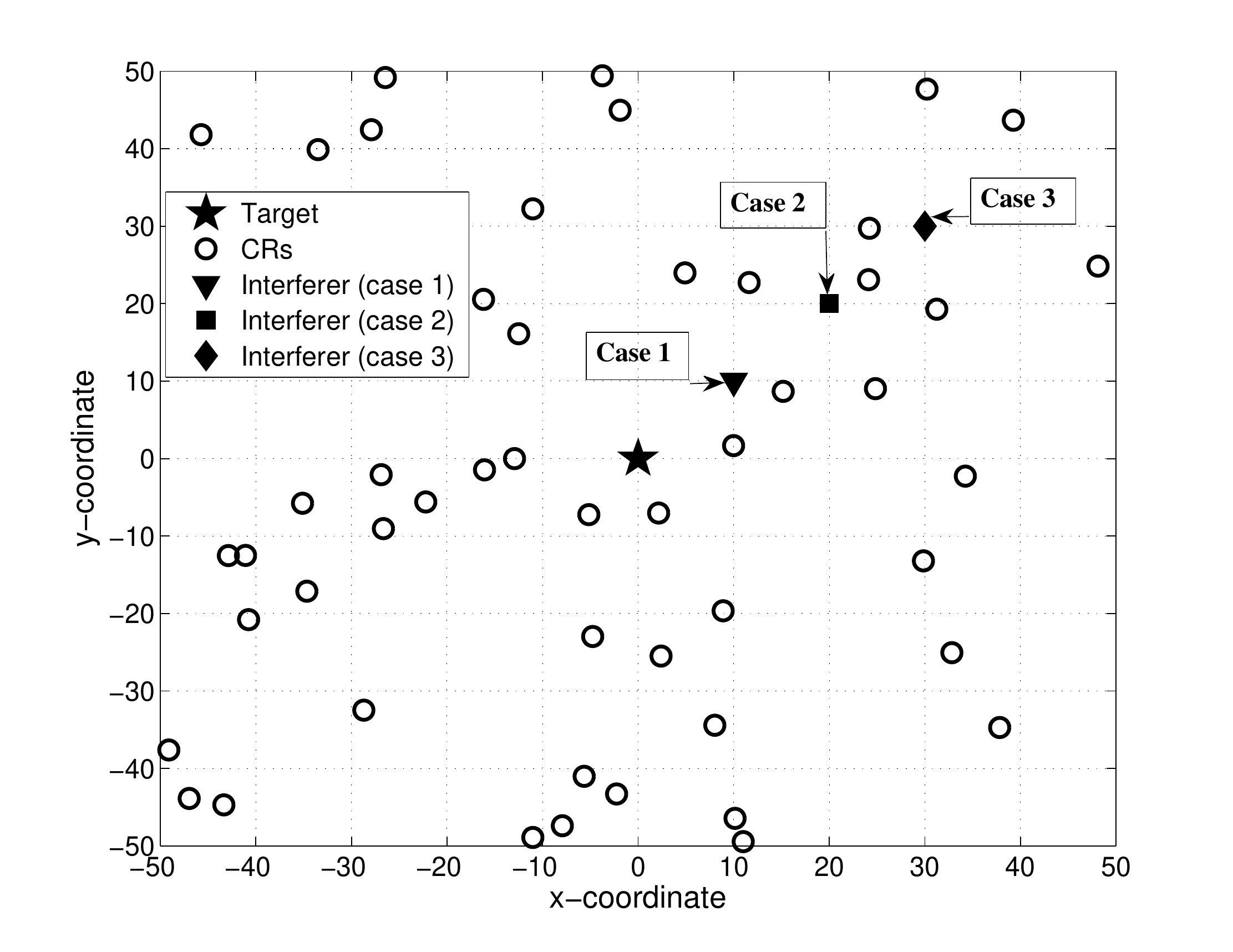}
\caption{Locations of the target, the interferer, and  CRs in the network. Case 1: Interferer @ [10,10], Case 2: Interferer @ [20,20], Case 3: Interferer @ [30,30].}
\label{fig:locations}
\end{figure}

We observe in Eqn. (\ref{eq:cac_r_expanded}) that the CAC of the received signal depends on the interferer power $p_i$. Therefore, as the power of the interference increases, the estimates of the target location shift towards the interferer location. Consider a CR network as depicted in Fig. ~\ref{fig:locations}, where the target is at [0,0]. We change the interferer position from [10,10] to [20,20] and then to [30,30] and compute RMSE in each case. 

\begin{figure}
\centering
\includegraphics[width=\columnwidth]{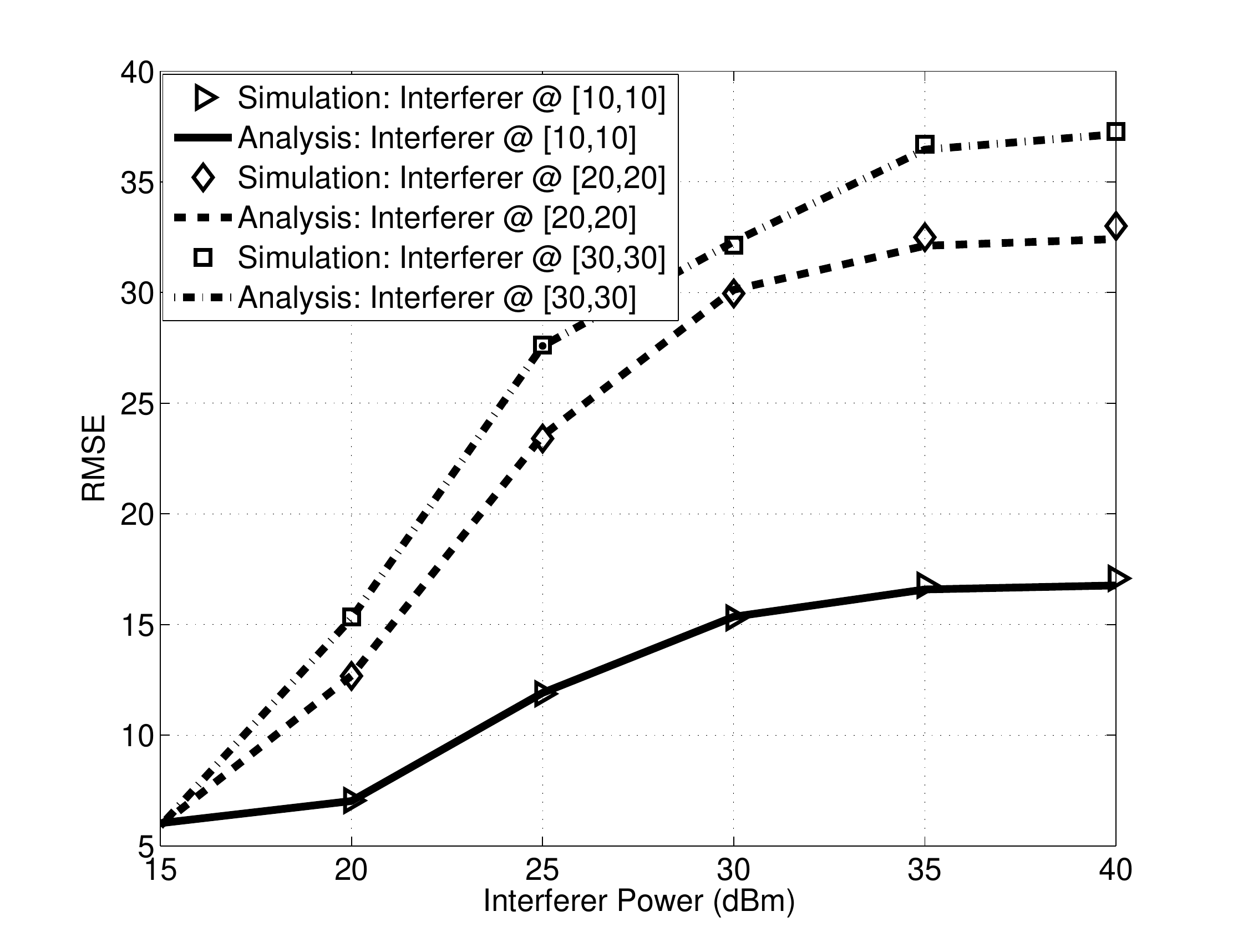}
\caption{RMSE for different locations of the interferer. Both the target and the interferer signals are 4-QAM. Target power $p_t$ = 10dBm.}
\label{fig:interferer_location}
\end{figure}

Consider one particular case when the target power, $p_t$ and the interferer power $p_i$ are 10 dBm and 30 dBm respectively. Now let us consider three experiments. In the first experiment, keep the interferer at [10,10] and find x-y coordinates of the target and hence the RMSE. Repeat this process for the second experiment with the interferer at [20,20] and the third experiment with the interferer at [30,30]. We observe that, as the interferer moves away from the origin (where the target is located), the estimates of the target location, too, move away from the origin. This results in higher RMSE. This is shown in Fig.~\ref{fig:interferer_location}. The RMSE curve for the second experiment is higher than that for the first and the RMSE curve for the third is higher than that of the first two experiments.

The phenomenon explained above, occurs because we are considering all the CRs in the network to find the location estimates. One way to improve the performance is to ignore CRs closer to the interferer. Identifying CRs closer to the interferer requires advanced processing and we do not address this issue in this paper.

Another important point to note from Fig.~\ref{fig:interferer_location} is that, all three RMSE curves tend to saturate when the interferer power is greater than 35 dBm. The RMSE value at which each curve saturates, depends on the distance between the target and the nearest CR to the interferer. In the first experiment (interferer @ [10,10]), this distance is approximately 17.48. We can see that saturation value of the RMSE curve is in the vicinity of this distance. Similar explanation can be extended for saturation value in the other two experiments.

\subsection{Modulation Scheme of the interferer}
\begin{figure}
\centering
\includegraphics[width=\columnwidth]{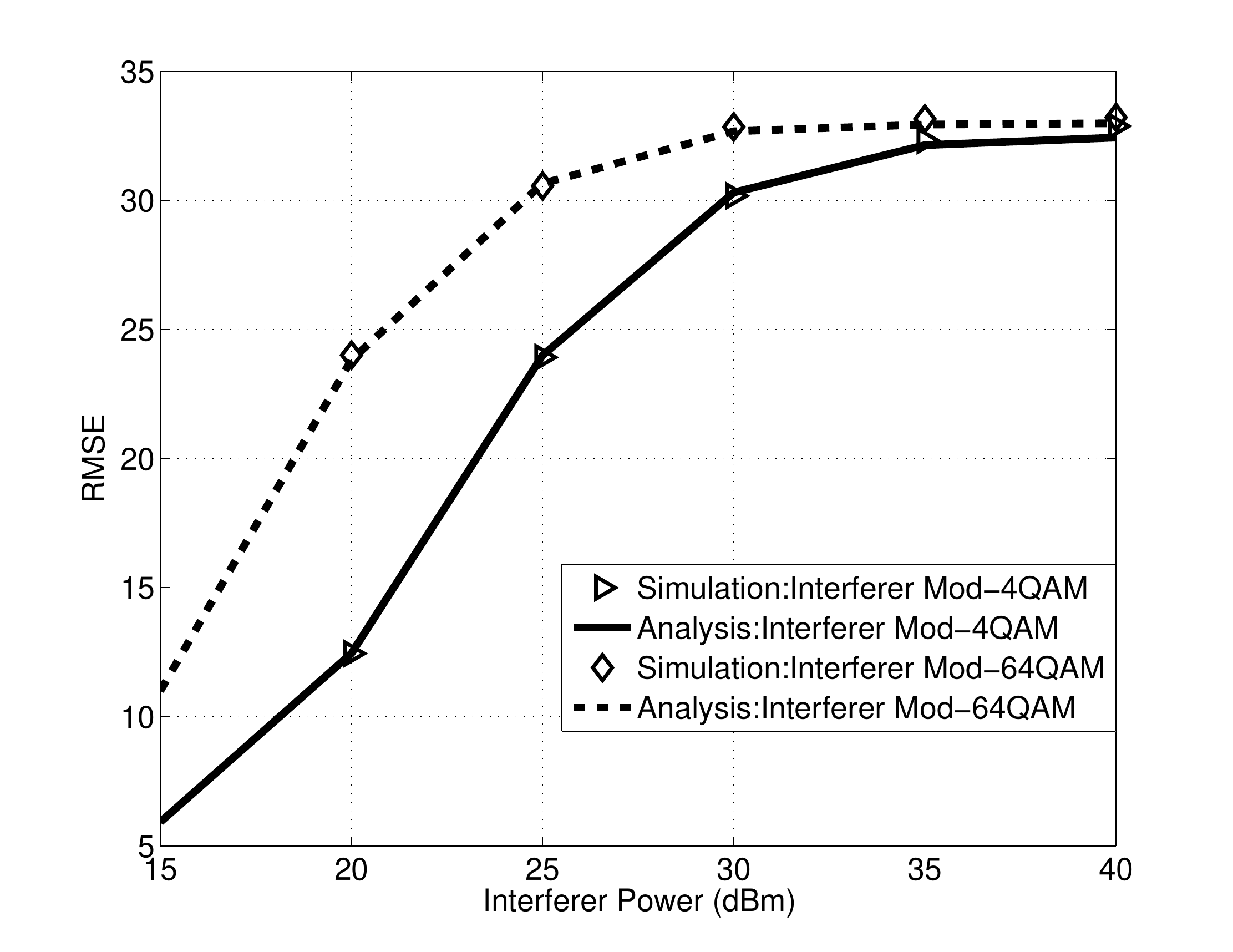}
\caption{RMSE for different modulation schemes of the interferer located at [20,20]. Target Modulation: 4-QAM, Target power $p_t$ = 10dBm.}
\label{fig:interferer_mod}
\end{figure}

In this section, we show how the modulation scheme of the interferer affects the RMSE performance. We see that the CAC of the received  signal (eq. \ref{eq:cac_r_expanded}) depends on the CAC of the interferer signal. If the modulation level of the interferer signal increases, the variations in the CAC of the interferer increases, which in turn results in higher variations in the CAC of the received signal. More variations in the CAC of the received signal lead to larger variance in the target location estimate $\hat x_t$ and $\hat y_t$, which increases RMSE. In Fig.~\ref{fig:interferer_mod}, we compare the RMSE when the interferer modulation is changed from 4-QAM to 64-QAM. Clearly higher modulation level gives rise to higher error.

However, it is important to note that, in both cases, the error tends to saturate at the same value as the interferer power is increased. We have seen in previous subsection, that the location estimates tends to be in the vicinity of the CR nearest to the interferer as the interferer power is boosted. In the present scenario, the interferer location is fixed at [20,20] in both cases. Therefore, the error tends to saturate at the same level.

\subsection{Impact of the target power level}
\begin{figure}
\centering
\includegraphics[width=\columnwidth]{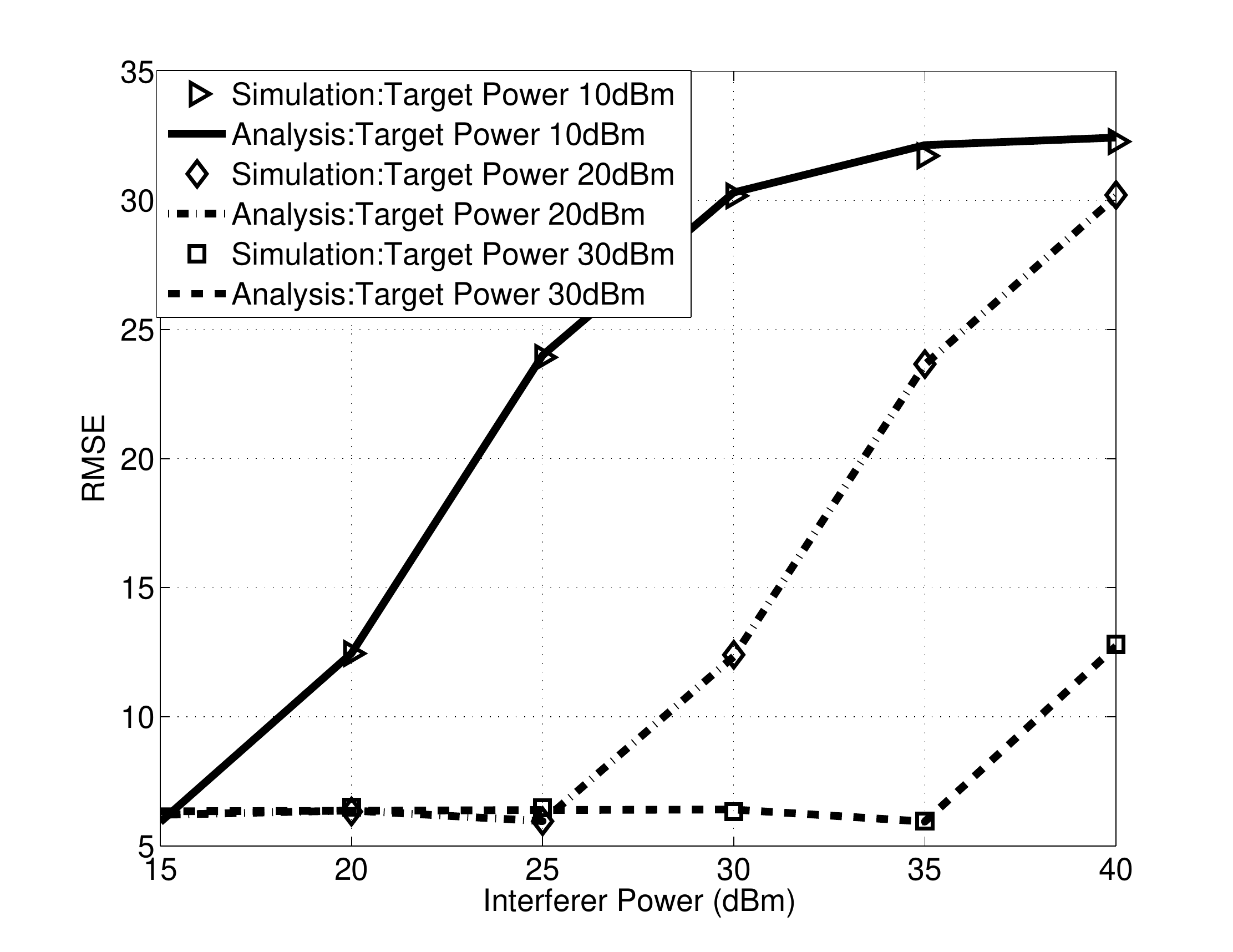}
\caption{RMSE for different target power levels. Interferer Location: [20,20]. The target and the interferer signals are 4-QAM.}
\label{fig:target_power}
\end{figure}

In weighted centroid algorithm, the estimates tend to be in the vicinity of the CRs that have higher value of the CAC of the received signal. As we increase the power transmitted by the target, the CRs located closer to the target get higher value of the CAC and hence higher weights, $\eta_k$. Therefore, the x-y coordinate estimates move towards these CRs and in turn, the estimates are closer to the actual location of the target. Therefore, higher target power results in lower RMSE as we can see in Fig.~\ref{fig:target_power}.

We can see, from the figure, that at a higher target power, the interferer needs to boost up its power to have any adverse impact on the location estimates. For example, when target power is fixed at 30dBm, the interferer power should be greater than 35dBm to increase the RMSE of the algorithm. If the interferer power is lower than 35dBm, the RMSE remains approximately constant.

\subsection{Comparison with WCL w/o cyclostationarity}
\begin{figure}
\centering
\includegraphics[width=\columnwidth]{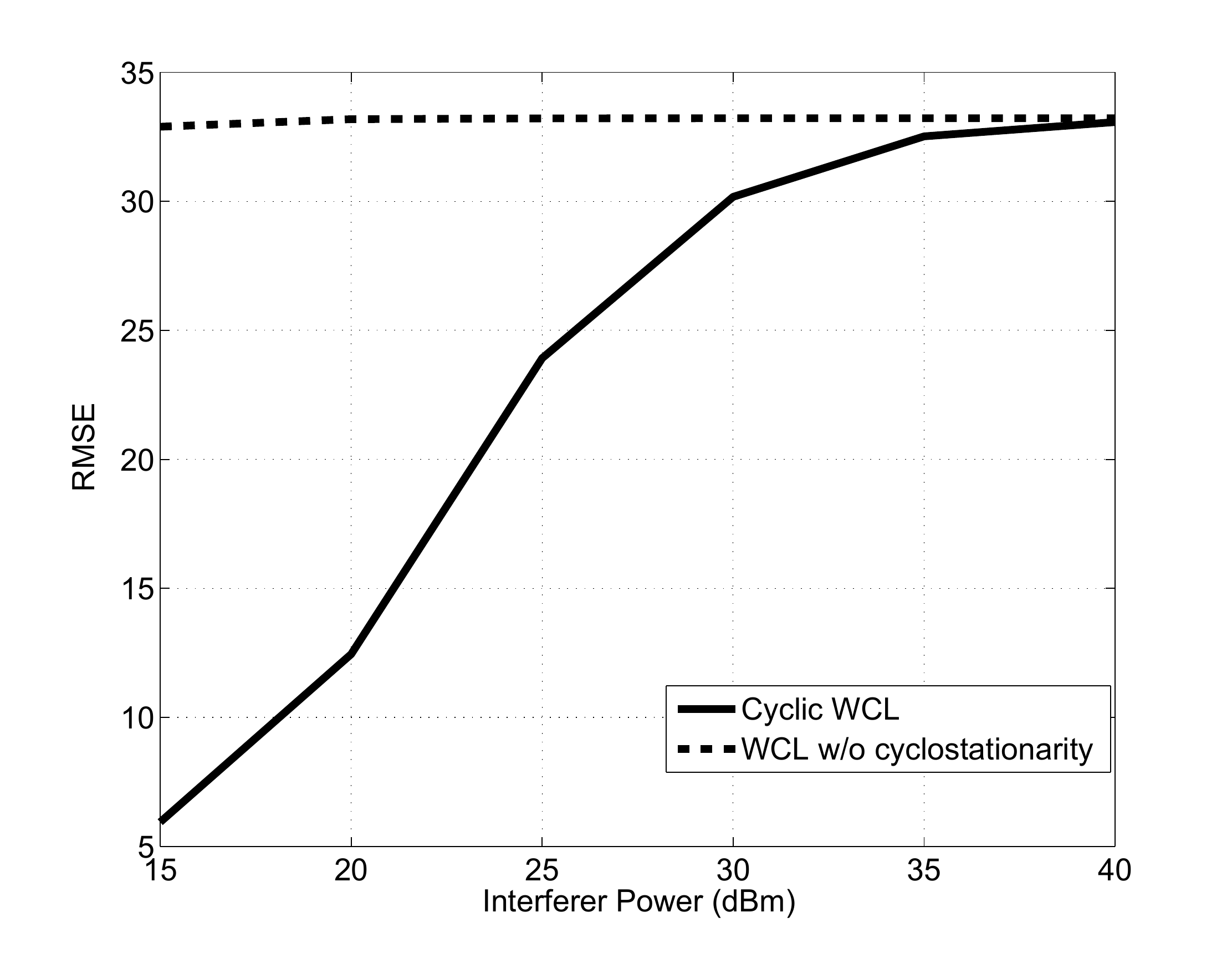}
\caption{Comparison between Cyclic WCL and WCL without cyclostationarity. Interferer Location: [20,20]. The target and the interferer signals are 4-QAM. Target power $p_t$ = 10dBm.}
\label{fig:wcl_cwcl}
\end{figure}

Now, we compare Cyclic WCL algorithm with WCL. WCL without cyclostationarity is simulated by setting cyclic frequency $\alpha_t$ to zero in the algorithm. As we can see in Fig.~\ref{fig:wcl_cwcl}, higher interferer power affects both WCL and Cyclic WCL. However, the error in WCL is higher by a factor of 5 as compared to Cyclic WCL when the interferer power is 15 dBm. This is because in the Cyclic WCL, we compute the weights based on the CAC at cyclic frequency of the target and the interferer does not share the same cyclic frequency. It is important to note that, even though the interferer does not have any cyclic feature at $\alpha_t$, it still causes some interference proportional to its power. This is because only a finite number of samples were used for CAC computation, which leads to a non-zero component of the CAC of the interferer signal at $\alpha_t$. Therefore $\hat R_{s_i}$, albeit small, is non-zero even though $s_i$ does not have cyclic frequency $\alpha_t$. This leads to the fact that the RMSE increases with higher interferer power in Fig.~\ref{fig:interferer_location}, \ref{fig:interferer_mod}, \ref{fig:target_power}, and \ref{fig:wcl_cwcl}.

\section{Conclusion}
\label{sec:Conclusion}
In this paper, we proposed Cyclic Weighted Centroid Localization algorithm, which estimates the location of a cyclostationary signal transmitter in a cognitive radio network in the presence of a spectrally overlapped interferer. This algorithm exploits the cyclostationary feature of the target signal in order to provide better estimates of the target location as compared to regular WCL without cyclostationarity.

We have formulated the target location estimates as ratios of quadratic forms in a Gaussian random vector. Using these ratios, we theoretically evaluate the performance of the proposed algorithm in terms of RMSE. Using simulations and theoretical analysis, we have shown that the RMSE performance of the proposed algorithm is affected by the location of the interferer, its modulation scheme and the transmitted power. Further we have also shown that, for the given locations of the CRs, the transmitter and the interferer, the Cyclic WCL algorithm performs better as compared to WCL without cyclostationarity when there is a spectrally overlapped interference is present in the network.


\appendix
\subsection{Proof: $\hat x_t$ is a ratio of quadratic form in $\boldsymbol{\hat \theta}$}
From (\ref{eq:xt}), we have 
\begin{equation}
\nonumber
{{\hat x}_t} = \frac{{{{\sum\limits_{k = 1}^K {\left\| {\left[ \begin{gathered}
  {\boldsymbol{\hat \theta_r }^T} \hfill \\
  {\boldsymbol{\hat \theta_i }^T} \hfill \\ 
\end{gathered}  \right]\boldsymbol{p_k}} \right\|} }^2}{x_k}}}
{{\sum\limits_{k = 1}^K {{{\left\| {\left[ \begin{gathered}
  {\boldsymbol{\hat \theta_r }^T} \hfill \\
  {\boldsymbol{\hat \theta_i }^T} \hfill \\ 
\end{gathered}  \right]\boldsymbol{p_k}} \right\|}^2}} }} = \frac{{\sum\limits_{k = 1}^K {\boldsymbol{p_k}^T\left[ {{\boldsymbol{\hat \theta_r }}{\text{ }}{\boldsymbol{\hat \theta_i }}} \right]} {\text{ }}{x_k}\left[ \begin{gathered}
  {\boldsymbol{\hat \theta_r }^T} \hfill \\
  {\boldsymbol{\hat \theta_i }^T} \hfill \\ 
\end{gathered}  \right]\boldsymbol{p_k}}}
{{\sum\limits_{k = 1}^K {\boldsymbol{p_k}^T\left[ {{\boldsymbol{\hat \theta_r }}{\text{ }}{\boldsymbol{\hat \theta_i }}} \right]} {\text{ }}\left[ \begin{gathered}
  {\boldsymbol{\hat \theta_r }^T} \hfill \\
  {\boldsymbol{\hat \theta_i }^T} \hfill \\ 
\end{gathered}  \right]\boldsymbol{p_k}}}
\end{equation}

\begin{equation}
\nonumber
\begin{gathered}
  {{\hat x}_t} = \frac{{Tr\left( {\left[ \begin{gathered}
  {\boldsymbol{\hat \theta_r }^T} \hfill \\
  {\boldsymbol{\hat \theta_i }^T} \hfill \\ 
\end{gathered}  \right]\mathop \sum \limits_{k = 1}^K \boldsymbol{p_k}{\text{ }}{x_k}\boldsymbol{p_k}^T\left[ {{\boldsymbol{\hat \theta_r}}{\text{ }}{\boldsymbol{\hat \theta_i }}} \right]} \right)}}
{{Tr\left( {\left[ \begin{gathered}
  {\boldsymbol{\hat \theta_r }^T} \hfill \\
  {\boldsymbol{\hat \theta_i }^T} \hfill \\ 
\end{gathered}  \right]\mathop \sum \limits_{k = 1}^K  \boldsymbol{p_k}\boldsymbol{p_k}^T\left[ {{\boldsymbol{\hat \theta_r}}{\text{ }}{\boldsymbol{\hat \theta_i }}} \right]} \right)}} \hfill \\
  {{\hat x}_t} = \frac{{\left[ {{\boldsymbol{\hat \theta_r }^T}{\text{ }}{\boldsymbol{\hat \theta_i }^T}} \right]\left[ {\begin{array}{*{20}{c}}
   {\boldsymbol{A_{p,x}}} & \boldsymbol{0}  \\
   \boldsymbol{0} & {\boldsymbol{A_{p,x}}}  \\

 \end{array} } \right]\left[ \begin{gathered}
  {\boldsymbol{\hat \theta_r}} \hfill \\
  {\boldsymbol{\hat \theta_i }} \hfill \\ 
\end{gathered}  \right]}}
{{\left[ {{\boldsymbol{\hat \theta_r }^T}{\text{ }}{\boldsymbol{\hat \theta_i }^T}} \right]\left[ {\begin{array}{*{20}{c}}
   {\boldsymbol{B_p}} & \boldsymbol{0}  \\
   \boldsymbol{0} & {\boldsymbol{B_p}}  \\

 \end{array} } \right]\left[ \begin{gathered}
  {\boldsymbol{\hat \theta_r}} \hfill \\
  {\boldsymbol{\hat \theta_i }} \hfill \\ 
\end{gathered}  \right]}} = \frac{{{\boldsymbol{\hat \theta }^T}\boldsymbol{A_x}\boldsymbol{\hat \theta} }}
{{{\boldsymbol{\hat \theta }^T}\boldsymbol{B}\boldsymbol{\hat \theta} }} \hfill \\ 
\end{gathered}
\end{equation}

where $\boldsymbol{A_{p,x}}$,$\boldsymbol{A_{x}}$, $\boldsymbol{B_{p}}$ and $\boldsymbol{B}$ are as defined in (\ref{eq:matrix_def}).

\subsection{Proof: $\boldsymbol{B_p}$ and $\boldsymbol{B}$ are positive definite matrices}
From definition of $\boldsymbol{B_p}$:
\begin{equation}
\nonumber
\boldsymbol{B_p} = \mathop \sum \limits_{k = 1}^K  \boldsymbol{p_k}\boldsymbol{p_k}^T = \left[ {\boldsymbol{p_1}{\text{ }}\boldsymbol{p_2}...\boldsymbol{p_K}} \right]\left[ \begin{gathered}
  \boldsymbol{p_1}^T \hfill \\
  \boldsymbol{p_2}^T \hfill \\
   \vdots  \hfill \\
  \boldsymbol{p_K}^T \hfill \\ 
\end{gathered}  \right] = \boldsymbol{P}\boldsymbol{P^T}
\end{equation}
where $\boldsymbol{P}=[\boldsymbol{p_{1}} {\text{ }} \boldsymbol{p_{2}}...\boldsymbol{p_{k}}]$. Now consider an arbitrary 3x1 vector $y$, then
\begin{equation}
\nonumber
\boldsymbol{y^T}\boldsymbol{B_p}\boldsymbol{y} = \boldsymbol{y^T}\boldsymbol{P}\boldsymbol{P^T}\boldsymbol{y} = {\left\| {\boldsymbol{P^T}\boldsymbol{y}} \right\|^2} \geqslant 0
\end{equation}

The last inequality comes from the fact that the square of 2-norm any vector is always non-negative. Therefore, $\boldsymbol{B_p}$ is always positive semi-definite.

Now let us consider the case where ${\left\| {\boldsymbol{P^T}\boldsymbol{y}} \right\|^2} \geqslant 0$:
\begin{equation}
\nonumber
\begin{gathered}
  {\left\| {\boldsymbol{P^T}\boldsymbol{y}} \right\|^2} = 0{\text{ }} \Leftrightarrow {\text{ }}\boldsymbol{P^T}\boldsymbol{y} = 0{\text{ }} \Leftrightarrow {\text{ Columns of }}\boldsymbol{P^T}{\text{ are }} \hfill \\
  {\text{ linearly dependent }} \Leftrightarrow {\text{ Rows of }}\boldsymbol{P}{\text{ are linearly dependent}}{\text{.}} \hfill \\ 
\end{gathered}
\end{equation}

But, 
 to definition of $\boldsymbol{P}$
\begin{equation}
\nonumber
\boldsymbol{P}{\text{  =  }}\left[ {\begin{array}{*{20}{c}}
   {{p_{t,1}}}  \\
   {{p_{i,1}}}  \\
   {\sqrt {{p_{t,1}}{p_{i,1}}} }  \\

 \end{array} \begin{array}{*{20}{c}}
   {{p_{t,2}}}  \\
   {{p_{i,2}}}  \\
   {\sqrt {{p_{t,2}}{p_{i,2}}} }  \\

 \end{array} {\text{  }}......\begin{array}{*{20}{c}}
   {{p_{t,K}}}  \\
   {{p_{i,K}}}  \\
   {\sqrt {{p_{t,K}}{p_{i,K}}} }  \\

 \end{array} } \right]
\end{equation}

Therefore, rows of $\boldsymbol{P}$ contain power received from the target (row 1), power received from the interferer (row 2) and square-root of product of these powers (row 3). Since the source and the interferer are located at different locations and the received power follows simplified path-loss model, clearly the rows are not linearly dependent. Therefore, we have

\[{\left\| {\boldsymbol{P^T}\boldsymbol{y}} \right\|^2} > 0{\text{ }} \Leftrightarrow \boldsymbol{B_p}{\text{ is positive definite}}{\text{.}}\]

Further, since $\boldsymbol{B} = diag(\boldsymbol{B_p},\boldsymbol{B_p})$, it is easy to see that $\boldsymbol{B}$ is also positive definite.

\bibliographystyle{IEEEtran}
\bibliography{references}

\end{document}